\documentclass[structabstract]{aa}
\usepackage{graphicx}
\usepackage{txfonts}
\usepackage{natbib}
\usepackage{hyperref}

\begin{document}

   \title{X-ray polarimetric signatures induced by spectral variability in the framework of the receding torus model}
   \titlerunning{Polarimetric signatures of the receding torus model}
   
   \author{F. Marin\inst{1} \thanks{\email{frederic.marin@astro.unistra.fr}}
          \and
           R.~W. Goosmann\inst{1}
           \and
           P.-O.~Petrucci\inst{2,}\inst{3}
	  }
   \authorrunning{Marin, Goosmann \& Petrucci}

   \institute{Observatoire Astronomique de Strasbourg, Universit\'e de Strasbourg,
              CNRS, UMR 7550, 11 rue de l'Universit\'e, 67000 Strasbourg, France
              \and Universit\'e Grenoble Alpes, IPAG, F-38000 Grenoble, France 
              \and CNRS, IPAG, F-38000 Grenoble, France }

   \date{Received 8 March, 2016; Accepted 26 April 2016}

  \abstract
  {Obscuring circumnuclear dust is a well-established constituent of active galactic nuclei (AGN). Traditionally 
  referred to as the receding dusty torus, its inner radius and angular extension should depend on the photo-ionizing 
  luminosity of the central source.}
  {We quantify the expected time-dependent near-infrared (IR), optical, ultraviolet (UV) and X-ray polarization of a 
  receding dusty torus as a function of the variable X-ray flux level and spectral shape.}
  {Using a Monte Carlo approach, we simulate the radiative transfer between the multiple components of an AGN adopting 
  model constraints from the bright Seyfert galaxy NGC~4151. We compare our model results to the observed near-IR 
  to UV polarization of the source and predict its X-ray polarization.}
  {We find that the 2 -- 8~keV polarization fraction of a standard AGN model varies from less then a few percent along 
  polar viewing angles up to tens of percent at equatorial inclinations. At viewing angles around the type-1/type-2 
  transition, there is a different X-ray polarization variability in a static or a receding torus scenario. In the former 
  case, the expected 2--8 keV polarization of NGC~4151 is found to be 1.21\% $\pm$ 0.34\% with a constant polarization 
  position angle, while in the latter scenario it varies from 0.1\% to 6\% depending on the photon index of the primary 
  radiation. Additionally, an orthogonal rotation of the polarization position angle with photon energy appears for very 
  soft primary spectra.}
  {Future X-ray polarimetry missions will be able to test whether the receding model is valid for Seyfert galaxies seen at 
  a viewing angle close to the torus horizon. The overall stability of the polarization position angle for photon indexes 
  softer than $\Gamma$ = 1.5 ensures that reliable measurements of X-ray polarization are possible. We derive a long-term 
  observational strategy for NGC~4151 assuming observations with a small to medium-sized X-ray polarimetry satellite.}

\keywords{Galaxies: active -- Galaxies: Seyfert -- Galaxies: individual : NGC 4151 -- Polarization -- Radiative transfer -- X-rays: galaxies}

\maketitle


\section{Introduction}
\label{Introduction}

Equatorial obscuration by an optically thick material that prevents near-infrared, optical, ultraviolet,
and soft X-ray radiation to escape from the subparsec regions of active galactic nuclei (AGN) is a mandatory 
element in our nowadays understanding of radio-quiet and radio-loud objects. There is evidence of circumnuclear
obscuration everywhere. The absence of optical broad lines, produced by Keplerian motion in a disk-like broad 
line region (BLR) in the spectra of type-2 (edge-on) AGN, only revealed by spectropolarimetry, was unambiguous 
proof of obscuration \citep{Antonucci1985}. X-ray observations provided additional confirmation of the presence 
of a Compton-thick molecular material close to the equatorial plane: soft ($<$ 2~keV) X-ray radiation from 
the power-law component were very much absorbed in type-2 Seyfert galaxies, while hard X-rays were ubiquitously 
detected for almost all objects \citep{Smith1996}. Their characteristic power-law emission spectrum, similar to 
that of type-1s but affected by photoelectric absorption cut-offs, showed that a cold molecular medium with a hydrogen 
column density up to 10$^{24}$~at/cm$^2$ (and even larger columns inferred for other sources) was indeed located 
between the source and the observer \citep{Turner1997}. 

The unified model, as postulated by \citet{Antonucci1985}, \citet{Lawrence1991}, \citet{Antonucci1993}, 
and \citet{Urry1995}, does not clearly define the morphology of the circumnuclear dust region. Originally treated as a 
toroidal structure, the advance of space observations reveals a more complex geometry, probably made of small optically 
thick clouds embedded in a radiation-driven outflow \citep{Elvis2000,Proga2007,Dorodnitsyn2015}. However, it was possible 
to estimate its outer radius (about 100 times the torus inner radius \citep{Kishimoto2009} and the very similar spectral 
energy distributions (SED) of AGN longward of 1$~\mu$m \citep{Elvis1994} points toward an inner radius fixed by dust 
sublimation. The inner radius is determined by the dust sublimation temperature and therefore depends on the luminosity 
of the photo-ionizing radiation originating in the central accretion disk. This suggestion first appeared in 
\citet{Lawrence1991}, where the concept of a ``receding torus'' was presented in agreement with observed time delays 
between different flux levels in the optical and near-IR continua \citep{Clavel1989,Oknyanskij2001,Koshida2009,Kishimoto2013}. 
Two additional strong arguments in favor of the receding torus scheme were made by \citet{Jackson1990}, \citet{Hill1996}, 
and \citet{Willott2000}: the radio-loud quasar fraction among radio-loud AGN increases with increasing 
emission-line and radio luminosities, and there is a difference in [O~{\sc iii}] line luminosity between radio galaxies 
and radio-loud quasars at a specific radio power. The same trends are detected in radio-quiet AGN 
\citep{Steffen2003,Kishimoto2013}, indicating that the inner radius of the equatorial structure must be influenced by 
the luminosity of the photo-ionizing source.

The luminosity affects the radius at which dust evaporates or condensates, and it is also  expected to 
affect the half-opening angle of the obscuring torus. \citet{Hill1996} have shown that the 3CR quasar fraction 
increases with increasing radio luminosity; however, this luminosity-size correlation seemed to be marginally 
significant at large ($z >$ 0.5) redshifts \citep{Kapahi1990}. \citet{Gopal1996} demonstrated that a wider opening 
angle with increasing radio luminosity could solve the problem, with the mean half-opening angle growing from about 
20$^\circ$ to 60$^\circ$ (with respect to the symmetry axis of the AGN defined by the direction of the radio jet) 
for objects with an [O~{\sc iii}] line luminosity ranging from 10$^{42}$ to 10$^{44}$ erg/s \citep{Arshakian2005}.
The fraction of type-2 AGN is thus expected to decrease with photo-ionizing luminosity \citep{Hill1996}, a 
trend that is confirmed by large samples of AGN observed in the optical \citep{Weymann1991,Simpson2005}, 
X-ray \citep{Fiore1998}, and radio bands \citep{Sanders1989}.

However, \citet{Willott2000,Willott2001} have shown that the luminosity-size correlation and the [O~{\sc iii}] line 
differences between radio galaxies and radio quasars can also be explained within reasonable margins by a two-population 
model with a unique torus at a fixed half-opening angle \citep{Grimes2004}. If this explanation is valid, the geometrical 
variability of a receding torus is not an absolute certitude in all quasar populations, and understanding the evolution of 
the morphology of the circumnuclear AGN material is crucial in order to assess the unified model. This need is highlighted by 
the discovery of strong absorption in the soft X-ray spectrum of NGC~5548, a famous Seyfert-1 with an estimated inclination of 
30$^\circ$ \citep{Ursini2015}, which contradicts the unified model. Similarly, a new but not well-quantified population of 
Seyfert-2 galaxies has appeared, showing little or no absorption in the X-ray band, and also a lack of broad optical lines 
\citep[e.g.,][]{Panessa2009,Bianchi2012}, a picture hardly compatible with an optically thick dusty torus situated close to 
the AGN equatorial plane. 

In this context, it is of a prime importance to investigate equatorial obscuring regions around AGN and to test whether changes 
in the torus properties can explain the observed particularities mentioned above. Since variations in torus half-opening angle 
and inner radius with photo-ionizing luminosity are expected to alter the angles at which photons will scatter on the molecular material,
a straightforward technique to achieve our goal is to look at the polarimetric signatures resulting from the receding torus model.
Hence, this paper aims to translate the different morphologies of the obscuring torus into detectable evidence in order to test the 
receding torus model. We focus our work on the X-ray polarimetric signatures of an AGN model based on NGC~4151 in order to assess 
scientific prospects for future X-ray polarimeters. We present the model and the Monte Carlo code used in this paper in Sect.~\ref{Model}, 
and we exploit the polarimetric results of different realizations of the torus in Sect.~\ref{Results}, with a special focus on NGC~4151 
in Sect.~\ref{NGC4151}. We discuss our results in the prism of the different timescales in the inner parsec of our model in 
Sect.~\ref{Discussion} and conclude in Sect.~\ref{Conclusions}.


\section{Modeling the receding torus}
\label{Model}

\subsection{Model setup and radiative transfer code}
\label{sec:model}

   \begin{figure}
   \centering
   \includegraphics[trim = 0mm 0mm 0mm 0mm, clip, width=9cm]{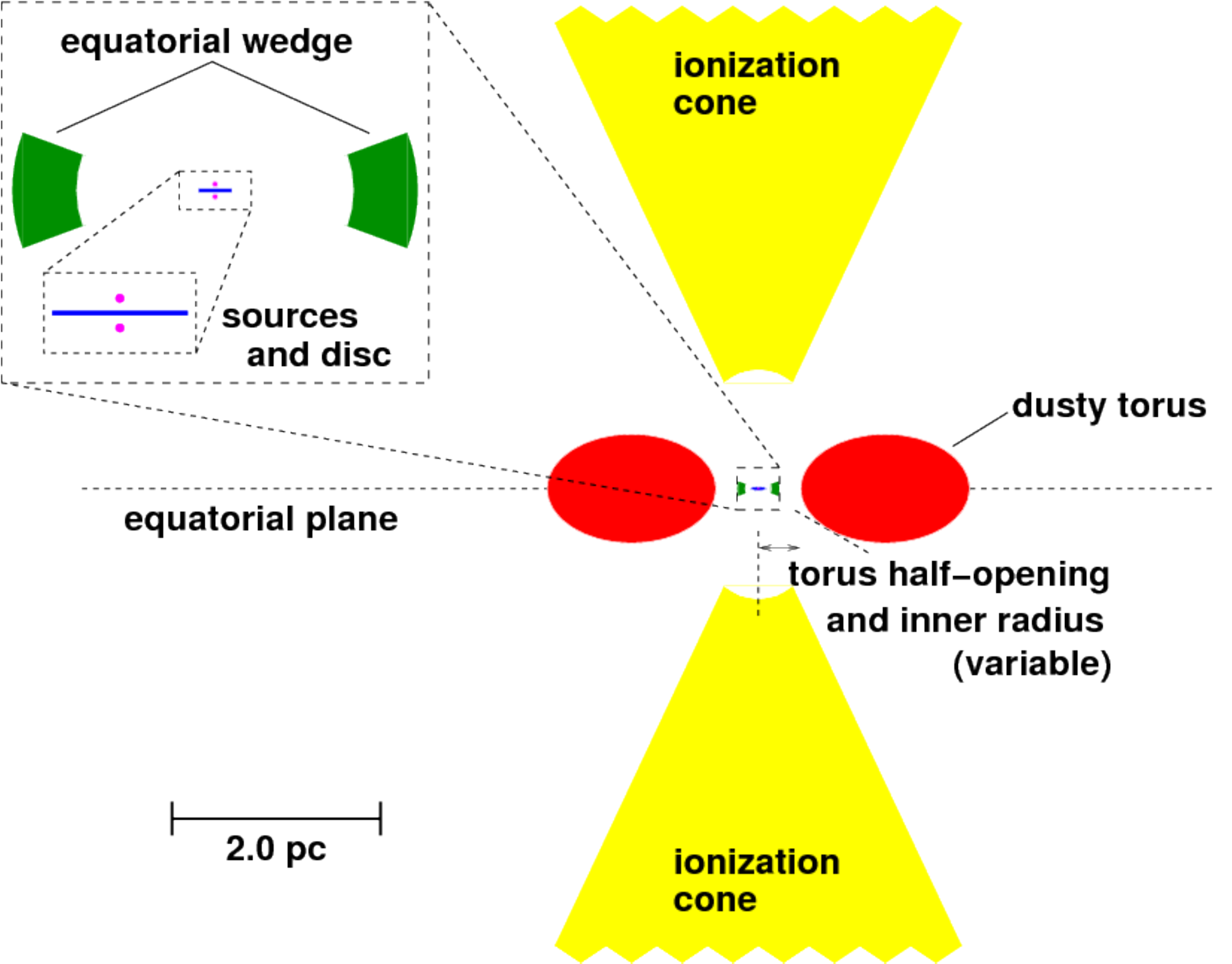}
      \caption{Emission and scattering geometry 
              in our baseline model. The box is a zoom on the central 
              AGN region. The outer radius of the polar outflows 
              (shown in yellow) are truncated to fit the sketch.
              Sizes and geometries are detailed in Table~\ref{Tab:Parameters}.}
     \label{Fig:Sketch}%
   \end{figure}

\begin{table*}[]
  \caption{Parametrization of the AGN model}            
  \centering
  {   
   \begin{tabular}{|l|l|}
   \hline {\bf Component}               & {\bf Description}\\
   \hline Continuum sources             & 2 point-like sources of isotropic emission,\\
          ~                             & power-law spectrum with a spectral flux $F_{\rm *}~\propto~\nu^{-\alpha}$,\\          
          ~                             & photon index $\Gamma$ = 1+$\alpha$ = variable,\\
          ~                             & distance from accretion disk mid-plane = $\pm$ 1 $\times$ 10$^{-3}$ pc\\          
   \hline Accretion disk                & cylindrical disk with $R_{\rm out}$ = 4 $\times$ 10$^{-3}$ pc,\\
          ~                             & half-thickness = 3.25 $\times$ 10$^{-7}$ pc,\\
          ~                             & irradiated by two isotropic sources,\\          
          ~                             & filled with optically thick cold matter\\          
   \hline Equatorial hot flow           & wedge structure with $R_{\rm in}$ = 3 $\times$ 10$^{-2}$ pc and $R_{\rm out}$ = 5 $\times$ 10$^{-2}$ pc,\\
          ~                             & half-opening angle (from the equatorial plane) = 20$^\circ$,\\          
          ~                             & equatorial Thomson optical depth = 1,\\  
          ~                             & filled with electrons\\               
   \hline Polar outflows                & double-cone with $R_{\rm in}$ = 1 pc and $R_{\rm out}$ = 400 pc,\\
          ~                             & half-opening angle (from the polar axis) = 33$^\circ$,\\          
          ~                             & vertical Thomson optical depth = 0.3,\\          
          ~                             & filled with electrons\\            
   \hline Dusty torus                   & elliptical torus with $R_{\rm in}$ = variable and $R_{\rm out}$ = 2.0 pc,\\
          ~                             & half-opening angle (from the polar axis) = variable,\\          
          ~                             & equatorial Mie optical depth $\gg$ 1,\\ 
          ~                             & filled with optically-thick cold matter\\            
   \hline
   \end{tabular}
  }
  \tablefoot{Parametrization of the AGN model used in this paper
           (references are given in the text). A sketch 
           of the model can be seen in Fig.~\ref{Fig:Sketch}.}
  \label{Tab:Parameters}
\end{table*}

The model we use in this paper follows the morphological parametrization of the Unification Scheme \citep{Antonucci1993}
and observational constraints on the composition and geometry of NGC~4151. At the center of the model is an optically thick,
geometrically thin accretion disk filled with mildly ionized matter. The disk reflects continuum photons originating 
from two hot plasma regions (coronas) above and below the disk. The coronas emit a power-law spectrum with a spectral flux 
that depends upon the photon index $\Gamma$: according to the known anti-correlation between the photon index and the Eddington 
ratio in Seyfert galaxies \citep{Gu2009}, increasing $\Gamma$ will result in decreasing X-ray fluxes. Along the equatorial plane 
lies a wedge scattering region that is responsible for the production of polarization parallel to the AGN radio axis at type-1 
inclinations, as detected in the ultraviolet (UV), optical, and near-infrared (NIR) regimes  \citep[e.g.,][]{Young2000,Smith2004}.
This ring-like disk is filled with electrons and has a small half-opening angle with respect to the equatorial plane.
Farther  out along the equator is the main obscuring region, an elliptical dusty torus with n$_{\rm H}$ = 10$^{25}$~at/cm$^2$. 
We fix its outer radius to 2~pc according to the most recent constraints obtained by \citet{Schnulle2015} on NGC~4151 using 
dust reverberation techniques. The inner radius of the torus and its half-opening angle (with respect to the polar 
axis) are left to vary according to the desired X-ray flux or $\Gamma$. Finally, along the pole, highly ionized winds  
carry out momentum and energy in the form of a double-conical region extending up to 400~pc \citep{Das2005}. Its half-opening 
angle (from the polar axis) is set to 33$^\circ$, as suggested by \citet{Das2005} based on long-slit spectroscopy and 
kinematic modeling of NGC~4151. A sketch of our multiple-component model is presented in Fig.~\ref{Fig:Sketch} and 
summarized in Table~\ref{Tab:Parameters}.

Radiative transfer will be undertaken by the Monte Carlo code {\sc stokes} \citep{Goosmann2007,Marin2012,Marin2015}, a multiple 
scattering code specifically designed to study the complex environment of Seyfert galaxies. The {\sc stokes} code computes the fluxes 
and polarization of radiation from the NIR to the hard X-ray band taking into account Mie, Thomson and Compton scattering, 
absorption, re-emission by fluorescent processes, photo-ionization, and recombination effects in a full three-dimensional 
environment. The Mie approximation breaks down at the X-ray energies considered here and should be replaced by 
calculations from X-ray diffraction theory (see \citealt{Draine2003}). This is not implemented in STOKES as the standard 
procedure for scattering on dust grains using Mueller matrices and Stokes formalism is not yet ready for X-rays \citep{Hoffman2016}. 
We therefore do not take dust scattering off of dust grains into account in the X-ray range, but rely only on the scattering 
component of the neutral reprocessing. So far, this is a rather common procedure in X-ray radiative transfer. Since the dust 
is optically thick, the polarization should be dominated by (single-)scattering at the surface of the medium, for which the 
electron scattering is dominant. The latest developments of {\sc stokes}, as well as a series of applications, are described 
in \citet{Marin2014}.

\subsection{Correlation between fluxes and torus morphology}
\label{sec:BSD}

   \begin{figure}
   \centering
   \includegraphics[trim = 0mm 0mm 0mm 0mm, clip, width=9cm]{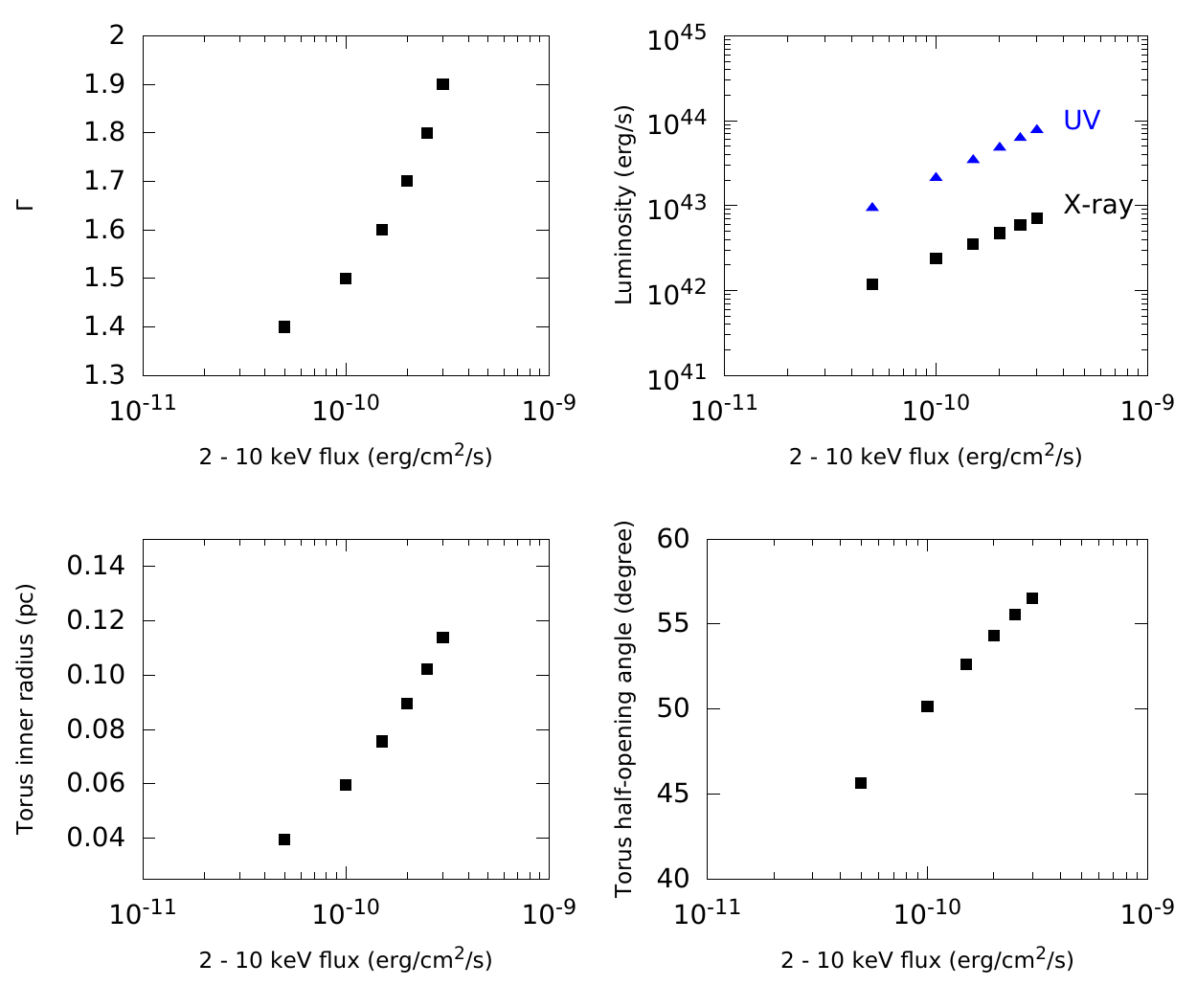}
      \caption{Variation in the key parameters of the torus 
               ($R_{\rm in}$ and half-opening angle), as well 
               as $\Gamma$ and the X-ray and UV luminosities of 
               the central engine, with respect to the initial 
               X-ray flux.}
     \label{Fig:Parametrization}%
   \end{figure}

The receding torus model developed by \citet{Simpson2005} and \citet{Arshakian2005} stipulates that the obscuring fraction 
of type-1 and type-2 AGN (either radio-quiet or radio-loud) depends on the luminosity of the central engine. In particular, 
\citet{Simpson2005} has shown that the height of the equatorial dusty torus varies with luminosity in Seyfert galaxies
as $\propto L^\alpha$, with $\alpha <$ 1. Similarly, \citet{Arshakian2005} found a statistically significant, positive 
correlation between the half-opening angle of the circumnuclear region of quasars and [O~{\sc iii}] emission-line luminosity. 
Hence, both the inner radius of the torus and its maximum height are luminosity dependent.

~\

From the conversion of the 2 -- 10~keV X-ray fluxes $F_{\rm X}$ to X-ray luminosities $L_{\rm X}$

   \begin{equation}
        L_{\rm X} = 4 \pi d^2 F_{\rm X} \label{Flux_Lum_X},
   \end{equation}where $d$ is the distance of the X-ray source in centimeters ($L_{\rm X}$ is in units of erg/s and $F_{\rm X}$ in erg/s/cm$^2$), it 
is possible to relate the change in $F_{\rm X}$ with UV luminosity $L_{\rm UV}$ using the relation found by \citet{Marchese2012}
when reconstructing the X-ray to optical SED of a sample of 195 X-ray selected type-1 AGN:   
   
   \begin{equation}
        L_{\rm UV} = 10^{1.18\log(L_{\rm X})-6.68} \label{Lum_X_UV}.
   \end{equation}

Once the UV luminosity (in erg/s) is estimated, it is easy to infer the inner boundary of the torus. If the torus inner wall
coincides with the dust sublimation radius, the estimation of R$_{\rm in}$ is given by \citet{Barvainis1987} and \citet{Kishimoto2007},
   
   \begin{equation}
        R_{\rm in} = 1.3 \left(\frac{L_{\rm UV}}{10^{46}}\right)^\frac{1}{2} \left(\frac{T_{\rm sub}}{1500}\right)^{-2.8} \left(\frac{a}{0.05}\right)^{-\frac{1}{2}} \label{Lum_UV_Rin},
   \end{equation}where T$_{\rm sub}$ is the dust sublimation temperature for graphite and silicate grains and $a$ is the grain radial size. 
\citet{Kishimoto2007} fixed T$_{\rm sub}$ to 1500~K, a value that accounts for the temperature range of ambient gas 
pressures for both graphite and silicate grains. The grain size was approximated by \citet{Barvainis1987} and fixed to 
0.05~$\mu$m in order to reflect the size distribution for interstellar graphite grains.

The torus height can be derived thanks to the empirical laws found by \citet{Simpson2005} and \citet{Arshakian2005},
who found a relation between the ionizing continuum luminosity and the torus maximum elevation, related to the 
type-1 AGN fraction $f$:
   
   \begin{equation}
        f = 1 - \left[1 + 3 \left(\frac{L_{\rm UV}}{L_0}\right)^{1 - 2\epsilon}\right]^{-\frac{1}{2}} \label{Lum_UV_H}.
   \end{equation}   
   Here $L_0$ is the ionizing luminosity at which the number of type-1 and type-2 objects is the same (so that the torus half-opening angle
is 60$^\circ$) and $\epsilon$ is an unknown exponent. Both values were numerically fitted to observational data by 
\citet{Simpson2005}, leading to a best-fit parametrization of $L_0$ = 10$^{34.39}$~W ($\sim$ 8 $\times$ 10$^{41}$~erg/s) and
$\epsilon$ = 0.23.

~\

Values for $\Gamma$, $L_{\rm X}$, $L_{\rm UV}$, $R_{\rm in}$, and the torus half-opening angle as a function of the initial X-ray flux 
are shown in Fig.~\ref{Fig:Parametrization}. The variation of $\Gamma$ with respect to $F_{\rm X}$ is taken from the Rossi X-Ray 
Timing Explorer monitoring of seven Seyfert 1 galaxies achieved by \citet{Markowitz2003}, with a special focus on NGC~4151.
Many authors have shown \citep[e.g.,][]{Nandra2001,Chiang2002,Singal2012} that the variation of $\Gamma$ with respect 
to $F_{\rm X}$  occurs on very short timescales ($<$ 2~ks, \citealt{Ponti2006}) and that  the photon index is usually softer for 
higher X-ray fluxes. In comparison to the kilo-second lags between $\Gamma$ and $F_{\rm X}$, the timescale needed for the torus to 
change its morphology is much longer ($\ge$ a year, see Sect.~\ref{Discussion:Timescales}). Therefore, hereafter we  consider that  $\Gamma$ and $F_{\rm X}$ vary instantaneously: fixing $\Gamma$ in our model is  equal to fixing 
$F_{\rm X}$, following the relation shown in Fig.~\ref{Fig:Parametrization} (top left). 

We note that the $L_{\rm UV}$ values derived from Eq.~\ref{Lum_X_UV} are consistent with the usual UV luminosities of NGC~4151 
($\sim$ 10$^{43 - 44}$ erg/s; see \citealt{Penston1990,Robinson1994,Cassidy1996,Minezaki2004}) and give rise to tori 
with inner radius  0.04~pc $< R_{\rm in} <$ 0.12~pc (Eq.~\ref{Lum_UV_Rin}). The half-opening angle of the circumnuclear 
structure (Eq.~\ref{Lum_UV_H}) varies between 45$^\circ$ and 57$^\circ$, in agreement with the receding torus scheme and 
the observational results of \citet{Arshakian2005}.


\section{Polarimetric results}
\label{Results}

The models investigated in this paper share a common root: NGC~4151. We selected this object for its high average X-ray and 
UV luminosities, which granting easier detection than  other less luminous AGN at the same redshift, but also because of  
the pre-existing collection of optical and infrared polarimetric observations that  allowed us to check the validity 
of our model. We note that the forthcoming results of our investigation are applicable to all Seyfert galaxies regardless of their 
inclination. Extending our results to slightly lower or higher luminosities is straightforward.

\subsection{UV, optical, and near-IR polarimetry}
\label{Results:Optical}

   \begin{figure}
   \centering
   \includegraphics[trim = 0mm 0mm 0mm 0mm, clip, width=9cm]{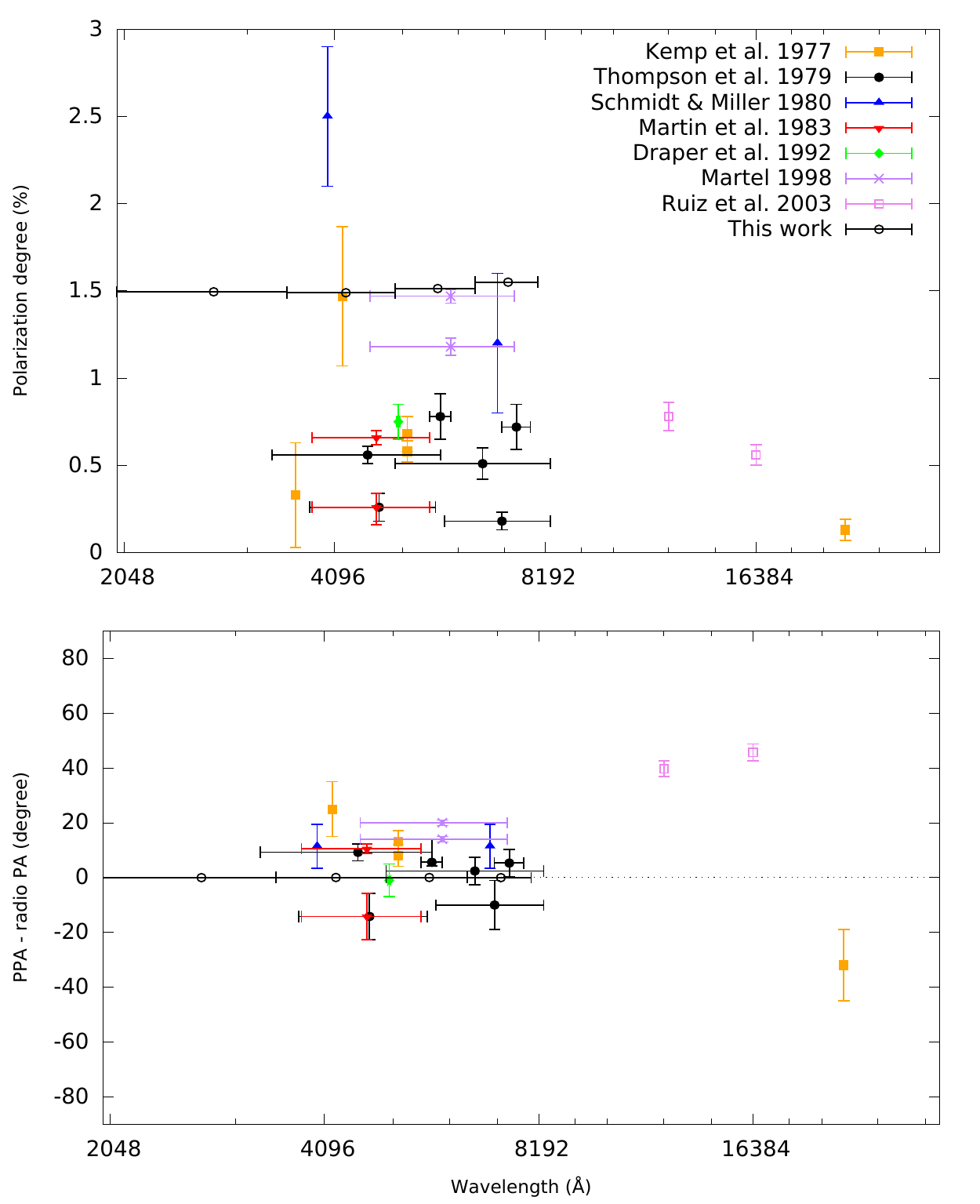}
      \caption{UV, optical, and NIR polarimetric measurements and results 
               of our baseline model based on NGC~4151. Top: polarization 
               degree $P$; bottom: difference between the radio position 
               angle $PA$ and the polarization position angle $PPA$
               (we use the convention $PPA$ = 0 for perpendicular 
               polarization angle and $PPA$ = 90 for parallel polarization).                
               Computational results from {\sc stokes}
               are shown by empty black circles.}
     \label{Fig:Optical}%
   \end{figure}

The {\sc stokes} code was simultaneously run in the UV, optical, and NIR bands in order to compare the polarimetric results of the 
code with published spectropolarimetric observations of NGC~4151. We sampled about 10$^{10}$ initially unpolarized photons, 
ranging from 2000 to 8000~\AA. The continuum photon index $\Gamma$ was set to 1.6, a value corrected for absorption averaged 
over 2.5 years \citep{Markowitz2003}. The resulting torus morphology is thus $R_{\rm in}$ = 0.076~pc, $R_{\rm out}$ = 2.0 pc, 
and half-opening angle = 52.64$^\circ$. We selected an observer's line of sight equal to 47$^\circ$ according to the modeling 
of \citet{Das2005} and \citet{Fischer2013}, who determined the inclinations of a sample of nearby AGN using their narrow line 
region (NLR) kinematics.

Results are shown in Fig.~\ref{Fig:Optical}, both for the linear continuum polarization degree $P$ and the polarization 
position angle ($PPA$). Our model produces a polarization degree on the order of 1.5\% in the UV and optical band, with a 
slight increase at larger wavelengths  due to the scattering phase function of the Milky Way dust model used in 
this simulation \citep{Mathis1977}. The computed $PPA$ is equal to 90$^\circ$ (a polarization angle parallel to the symmetry 
axis of the model), the typical signature of type-1 AGN.

Comparing our results with archival broadband polarimetric observations, we find that our model successfully reproduces 
both the polarization degree and position angle of literature data. In particular, the polarization degree resulting from
the code is in agreement with the measures of \citet{Kemp1977} despite the large variation in $P$ found by the author
around 4200\AA. At this wavelength, the telescope aperture was 15~arcsec, while at larger wavelengths the aperture decreased 
 to 5.9~arcsec. The contribution of the extended ($\sim$ 400~pc) NLR, where perpendicular scattering occurs, is then less
important and the resulting net polarization is lower. Our simulation also agrees with the 7000~\AA~polarization measurements done 
by \citet{Schmidt1980} and the optical observations of \citet{Martel1998}. However, the optical polarimetry achieved by 
\citet{Thompson1979}, \citet{Martin1983}, and \citet{Draper1992} appears to have smaller $P$. Similarly to \citet{Kemp1977},
the authors focused on the core polarization (using apertures smaller than 5~arcsec), neglecting a large fraction of 
the NLR-scattered polarization. The mid-IR polarimetric observations achieved by \citet{Ruiz2003} indicates that $P$
should decrease with increasing wavelength as a result of the dominance of unpolarized dust reemission over Thomson scattering.
Overall, all the polarimetric observations find a polarization position angle coherent with equatorial scattering dominance:
the difference between the position angle of the radio axis (77$^\circ$, as measured by \citet{Mundell2003} and references
therein) and the $PPA$ is $\sim$ 0$^\circ$.

\subsection{X-ray polarimetry}
\label{Results:X}

We computed the integrated 2 -- 8~keV polarization from our model for 20 viewing angles $\theta$ equally distributed in 
cos($\theta$). Dilution from fluorescence emission was taken into account for all the principal elements in the table of 
abundances of \citet{Anders1982}, with most of the unpolarized line photons emitted by the 6.4~keV iron line. The specific 
energy range of this analysis was  selected in order to allow for future comparison with the forthcoming generation of 
X-ray polarimeters (see Sect.~\ref{Discussion:Polarimeters}).

Our analysis is divided into two steps. First, we  investigate an X-ray model where the torus is insensitive to 
the radiation field, i.e., its inner radius and half-opening angle remain constant with variable $\Gamma$/$F_{\rm X}$,
so we can check what  the expected X-ray polarization is when the torus has no time to react to 
the variation in the radiation field. Second, we allow the torus to vary with $\Gamma$/$F_{\rm X}$, 
considering that the timescale of dust destruction/formation by heating/cooling mechanisms is on the same order 
as the X-ray variability. The real timescale for the torus morphology to react to the change in the central source 
intensity is probably between the two assumptions and will be discussed in Sect.~\ref{Discussion:Timescales}.

\subsubsection{The static torus case}
\label{Results:X:Stability}

   \begin{figure}
   \centering
   \includegraphics[trim = 10mm 0mm 8mm 8mm, clip, width=9cm]{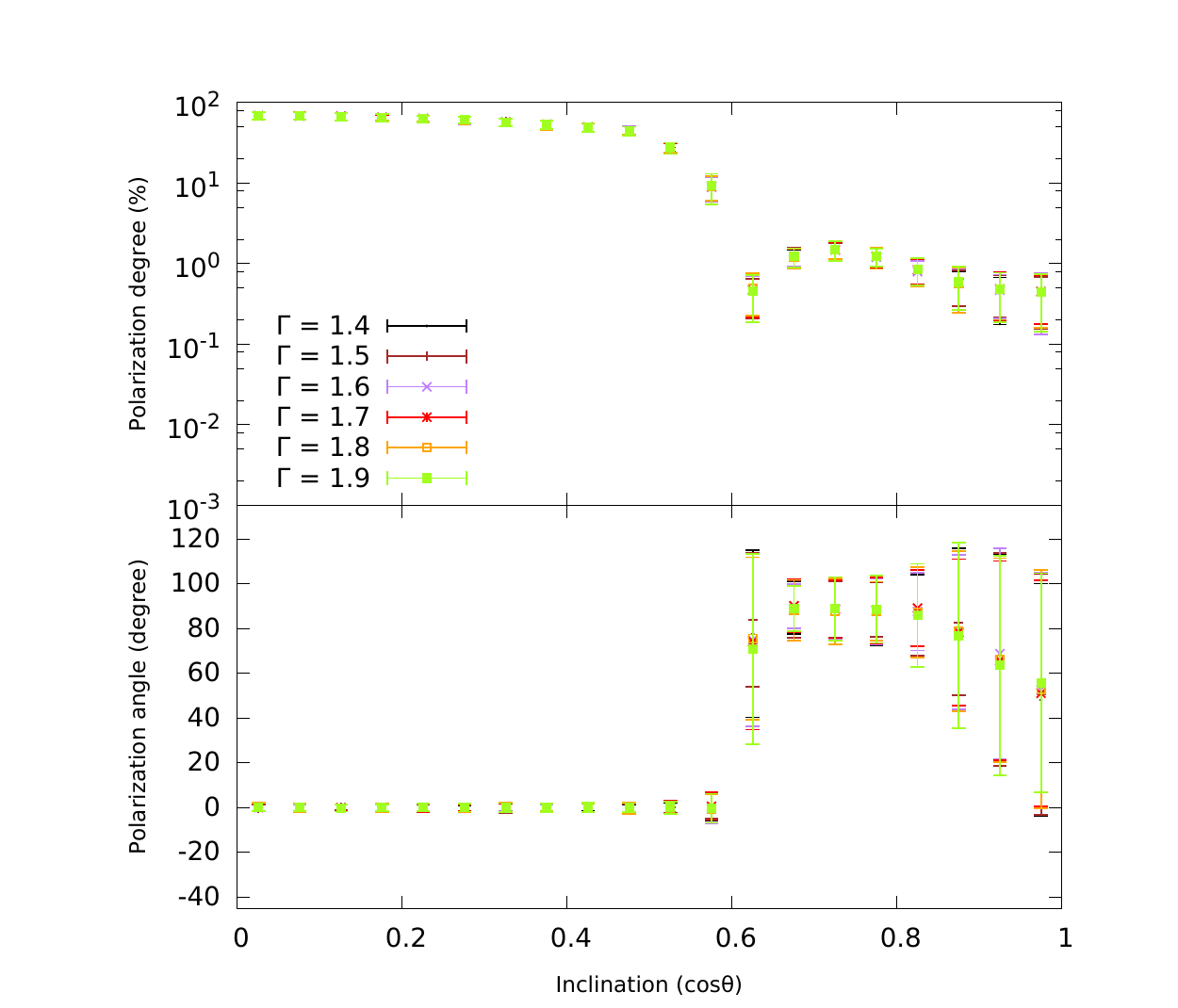}
      \caption{Integrated 2 -- 8~keV polarization degree (top)
               and polarization position angle (bottom) of our 
               baseline model as a function of the observer's 
               viewing angle. In this particular case, the torus
               morphology is considered to have no time to 
               react to the variation in the intensity of 
               the radiation field.}
     \label{Fig:X_stability_angles}%
   \end{figure}

We fix the internal torus radius to 0.076~pc and its half-opening angle to 52.64$^\circ$ with respect to the 
polar axis. This parametrization is thus the same as the NIR/optical/UV model investigated above. We allow 
$\Gamma$ to vary from 1.4 to 1.9 and look at the integrated 2 -- 8~keV polarization. Results are show in 
Fig.~\ref{Fig:X_stability_angles}. A direct result is that $P$ is almost insensitive to the variation in 
$\Gamma$ (or $F_{\rm X}$), the difference being lost in the polarization statistics. The dominance of 
photo-absorption and line emission over scattering in the soft X-ray band explains that the polarization 
fraction is not very sensitive to the hardness of the initial spectrum.

The integrated polarization degree resulting from multiple scattering is found to be below 2\% at 
type-1 inclinations with a variation that traces the nuclear inclination. At viewing angles closes to 
the torus horizon, parallel polarization emerging from the accretion disk and the equatorial structure 
compete with perpendicular scattering from the polar regions and the resulting $P$ decreases. When polar 
scattering finally dominates, the polarization fraction increases to 70\%. The dichotomy between type-1 
and type-2 Seyfert galaxies is clearly visible in the $PPA$ plot, where the transition is highlighted by 
a 90$^\circ$ rotation of the polarization angle. This phenomenon is analogous to what is observed in the 
optical band \citep{Antonucci1993}.

\subsubsection{ Receding torus case}
\label{Results:X:Variation}   
   
   \begin{figure}
   \centering
   \includegraphics[trim = 10mm 0mm 8mm 8mm, clip, width=9cm]{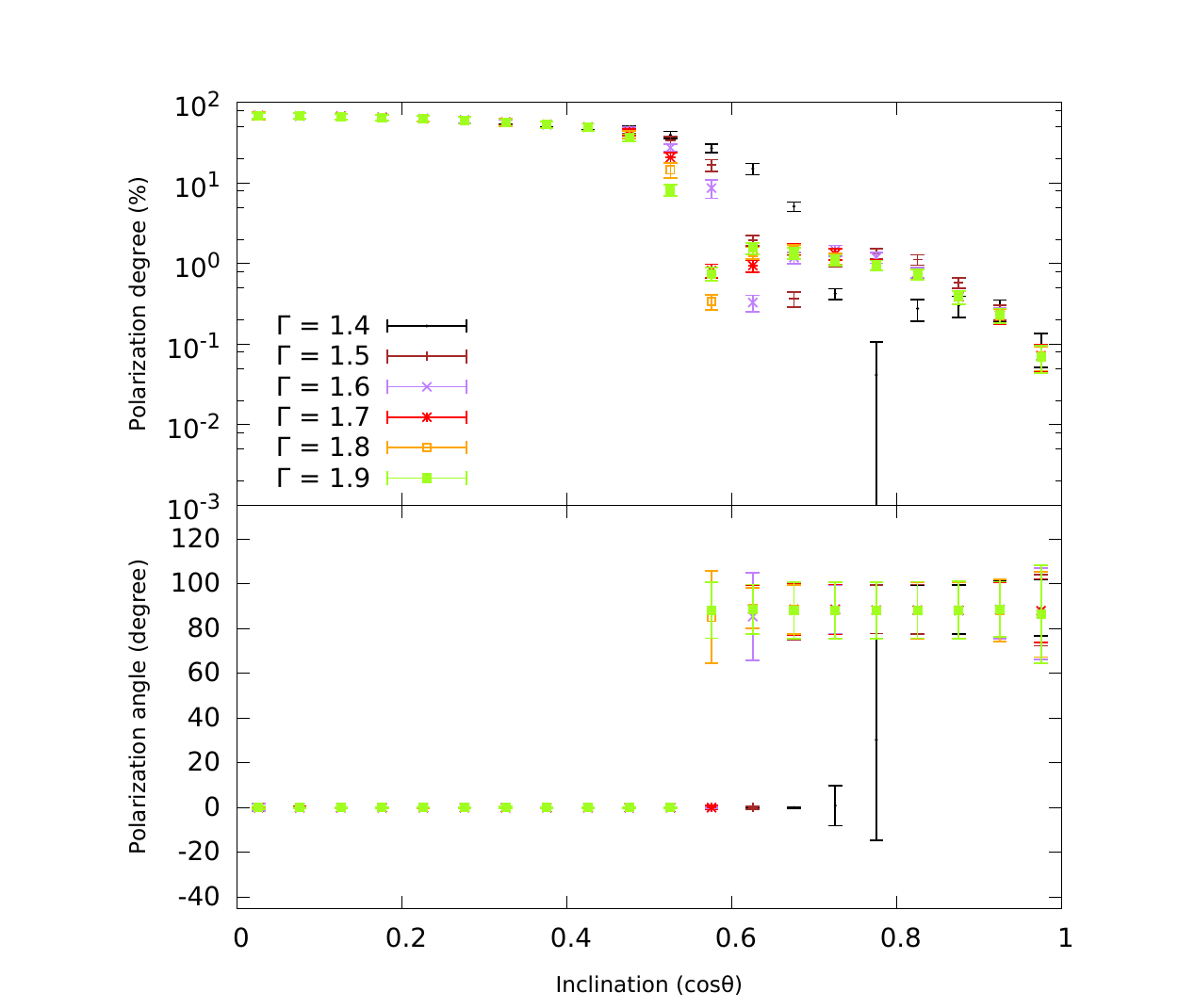}
      \caption{Integrated 2 -- 8~keV polarization degree (top)
               and polarization position angle (bottom) of our 
               baseline model as a function of the observer's 
               viewing angle. The torus morphology is now 
               dependent on the variation in radiation intensity.}
     \label{Fig:X_variation_angles}%
   \end{figure}   
   
We now allow the torus to vary instantaneously with the change in the spectral slope according to the results
presented in Fig.~\ref{Fig:Parametrization}. The  {\sc stokes} code, which treats static models, is not able to handle 
slow physical variations of its morphological parametrization with respect to the radiation field. We thus ran 
a series of ``snapshots'' because each model is parametrized with a different photon index. Despite
the variation in torus $R_{\rm in}$ and half-opening angle, we note that the half-opening angle of the NLR region remains set 
to a value measured by \citet{Das2005}; we thus ensure that our model  focuses solely on the polarimetric variations 
due to the receding torus scheme. Slightly higher NLR half-opening angles are not expected to strongly affect 
the final polarization \citep{Marin2012,Marin2015}.

The results of our modeling are compiled in Fig.~\ref{Fig:X_variation_angles}. The situation is quite different 
from the static torus case where the integrated 2 -- 8 keV polarization is strongly dependent on the 
torus half-opening angle, which indirectly translate into $\Gamma$/$F_{\rm X}$. For harder initial spectra 
(small $\Gamma$, bulky tori), the transition between type-1 and type-2 classifications occurs at smaller 
inclinations and the resulting polarization degree is quite low ($<$ 1\%) over a wide $\theta$ range (0 -- 45$^\circ$). 
With softer $\Gamma$, the transition occurs at larger inclinations owing to the variation in the torus height. It results 
in a blend of the polarimetric signal between 40$^\circ$ and 60$^\circ$ where the same object could be detected as a 
type-1 or a type-2 AGN depending on its initial $\Gamma$/$F_{\rm X}$. This is the typical signature of ``changing-look'' 
AGN, which are bright type-1 AGN known to show extreme variations (from Compton-thin to Compton-thick regimes; see, e.g., 
\citealt{Risaliti2005,Puccetti2007,Marchese2012}) in their hydrogen column density situated along the observer's 
line of sight\footnote{The timescales for Compton-thin to Compton-thick variation in n$_{\rm H}$ is object dependent. 
In the case of NGC~4151, the regime transition happens in timescales of years; see Sect.~\ref{NGC4151}.}. At larger 
inclinations, there is no impact of $\Gamma$/$F_{\rm X}$ to be detected in the X-ray polarization signal.

~\

According to our simulations, the expected X-ray polarization degree of type-1 Seyfert galaxies should be lower
than a few percent, while at type-2 inclinations, $P$ might rise to several tens of percent if there is 
no source of dilution nearby. This is of course an ideal case as scattered light will probably not be the dominant 
component of the observed flux. Unpolarized light, which dilutes the net polarization degree, might arise from circumnuclear
starburst activity and from a multitude of faint sources such as accreting white dwarfs and coronally active stars
\citep{Revnivtsev2009,Marin2015b}. Nevertheless, the levels of polarization predicted by this model are in agreement 
with the predictions for NGC~1068 achieved by \citet{Goosmann2011}. These predictions are not affected by the receding
torus model, which only affects intermediate AGN types.


\section{Application to NGC~4151}
\label{NGC4151}

NGC~4151 is a spiral galaxy\footnote{Apparently dubbed the ``Eye of Sauron'' by the astronomy community.} 
harboring a type-1 AGN situated at a distance of 14.066 $\pm$ 8.694~Mpc (mean distance and standard deviation 
for nine redshift-independent distances obtained with AGN time-lag and Tully-Fisher methods, according to the 
NASA/IPAC extragalactic database). As previously stated, NGC~4151 stands out thanks to its high X-ray fluxes 
and proximity, turning it into an ideal target for ground-based and space-borne polarimetric observations. NGC~4151 
is also know to be a changing-look AGN, where the observed column density of neutral gas varied from 
$\sim$ 2 $\times$ 10$^{23}$~at/cm$^{-2}$ in 1996 July to $\sim$ 9 $\times$ 10$^{22}$~at/cm$^{-2}$ in 2001 December \citep{Puccetti2007}. 
The changing-look nature of NGC~4151 is in agreement with the nuclear inclination (45$^\circ$ $\pm$ 5$^\circ$) 
derived by \citet{Das2005} and \citet{Fischer2013} and, as seen in Fig.~\ref{Fig:X_variation_angles}, the column 
density variation could be easily explained by a change in the structure of the obscuring region.

We now   study in greater detail the expected polarimetric signature of NGC~4151 in the framework of the 
static and receding torus schemes. To this end, we took the observed 2 -- 10~keV light curves for NGC~4151 
continuum flux and photon index from the Rossi X-Ray Timing Explorer monitoring of \citet{Markowitz2003}. The 
flux and $\Gamma$ variability will be used as initial inputs for our model. As  in Sect.~\ref{Results}, 
we alternatively investigate the static and receding torus models for a fixed observer's viewing angle.

\subsection{ Static torus case}
\label{NGC4151:Stability}

   \begin{figure}
   \centering
   \includegraphics[trim = 5mm 10mm 0mm 10mm, clip, width=9.8cm]{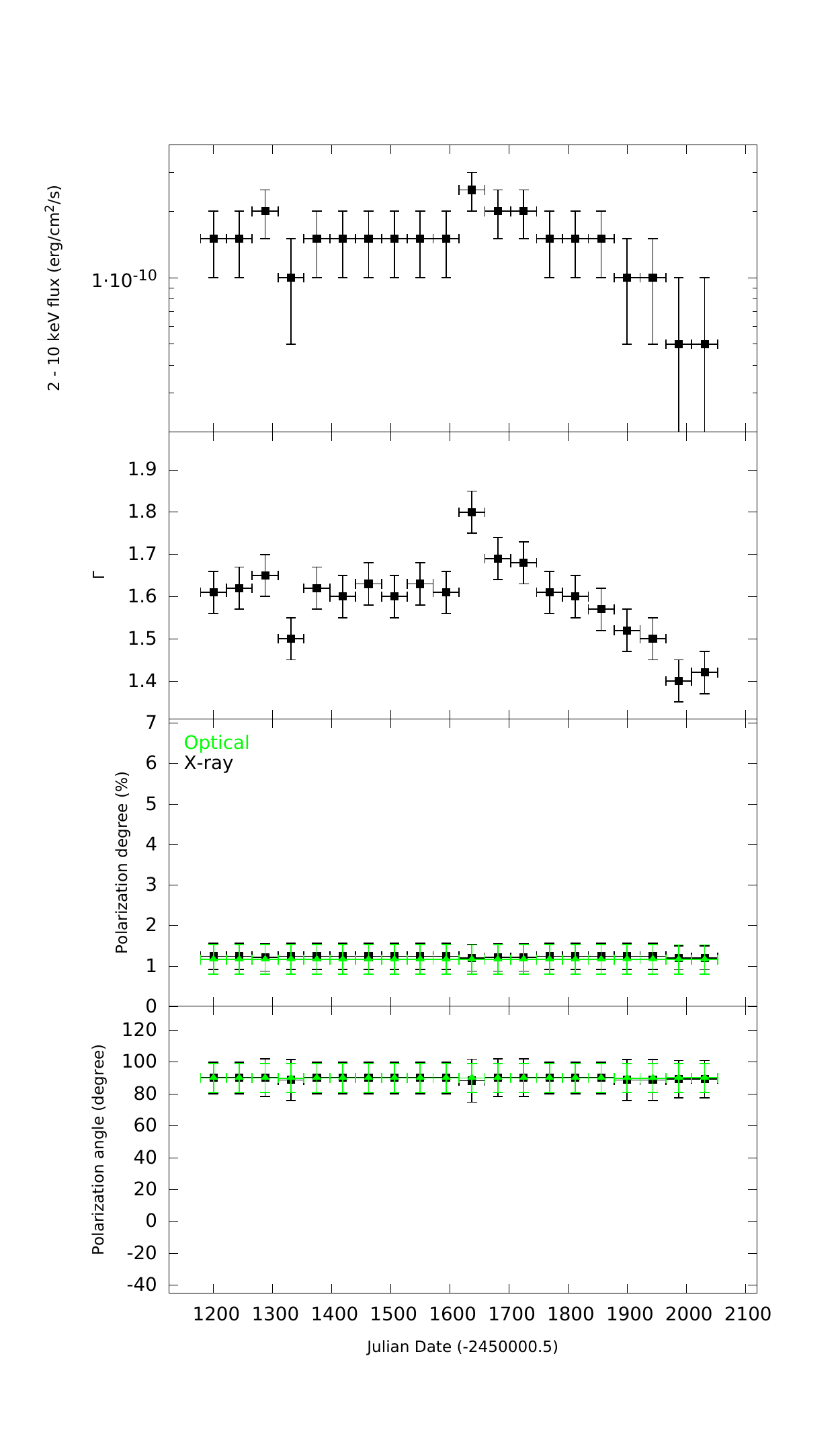}
      \caption{Long timescale light curves for continuum 2 -- 10~keV
               flux, $\Gamma$, polarization degree, and polarization 
               position angle (optical polarization in green, 
               X-ray polarization in black) for a static NGC~4151 model. The 
               continuum flux and photon index light curves are taken 
               from real observations by \citet{Markowitz2003}. 
               Consecutive observations are separated by $\sim$ 30~days.}
     \label{Fig:NGC_4151_stability}%
   \end{figure}   
   
The first scenario, which implies a torus morphology that does not have time to react to the change in the initial 
spectral slope, is presented in Fig.~\ref{Fig:NGC_4151_stability}. The observer's line of sight is not obscured 
by the torus horizon and the central engine is directly visible, so despite an important variation in $\Gamma$
(from 1.4 to 1.9), the computed polarization signal shows very little change. We show both the optical and X-ray
polarization resulting from the model, and find that the expected polarization degree of NGC~4151 is 1.21\% $\pm$ 0.34\% 
in the X-ray band with a constant parallel polarization position angle. The optical continuum polarization degree 
and polarization angle are very similar to the X-ray values as the dominant mechanism for scattering-induced polarization 
is electron scattering. The absorbing medium, either dust grains or atoms, efficiently block radiation along the equatorial 
plane in the 2 - 10~keV and optical bands. Differences can arise at smaller wavelengths where hard X-rays can penetrate 
deeper into the cold gas and escape from the circumnuclear medium \citet{Smith1996}, but at these two wavebands X-ray 
and optical polarization levels are  similar. The variability in fractional polarization for transitions 
between $\Gamma$ = 1.9 and $\Gamma$ = 1.4 is not detectable and lost in the error bars. 
  
The stability of the polarization degree and the polarization position angle despite the variation in intrinsic 
X-ray flux indicates that a future X-ray polarimetric observation would not be affected by variation of the mass 
accretion rate at the observed level. This would ensure the polarimetric observation to be reliable on long 
timescales, without the potential effect of polarization angle rotations (such as discussed in the next section).
Even though we assumed the timescales for dust formation and destruction are longer than the flux variation, we now 
relax this assumption.

\subsection{ Receding torus case}
\label{NGC4151:Variation}   

   \begin{figure}
   \centering
   \includegraphics[trim = 5mm 10mm 0mm 10mm, clip, width=9.8cm]{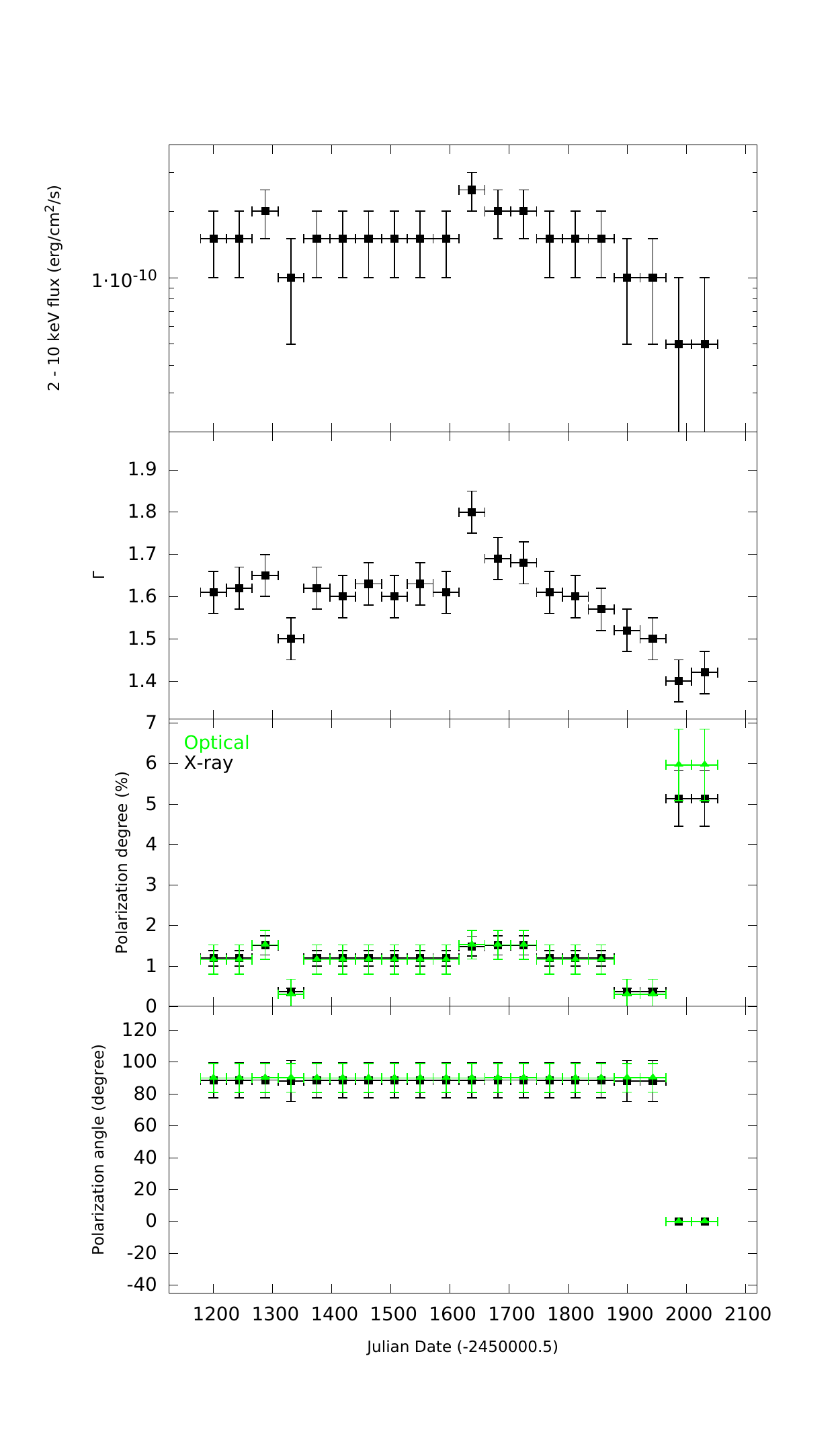}
      \caption{Long timescale light curves for continuum 2 -- 10~keV
               flux, $\Gamma$, polarization degree, and polarization 
               position angle (optical polarization in green, 
               X-ray polarization in black) of NGC~4151 in the case of a 
               $\Gamma$-sensitive torus. The continuum flux and photon 
               index light curves are taken from real observations by 
               \citet{Markowitz2003}. Consecutive observations 
               are separated by $\sim$ 30~days.}
     \label{Fig:NGC_4151_variation}%
   \end{figure}   

\begin{figure}
   \centering
   \includegraphics[trim = 10mm 0mm 8mm 8mm, clip, width=9cm]{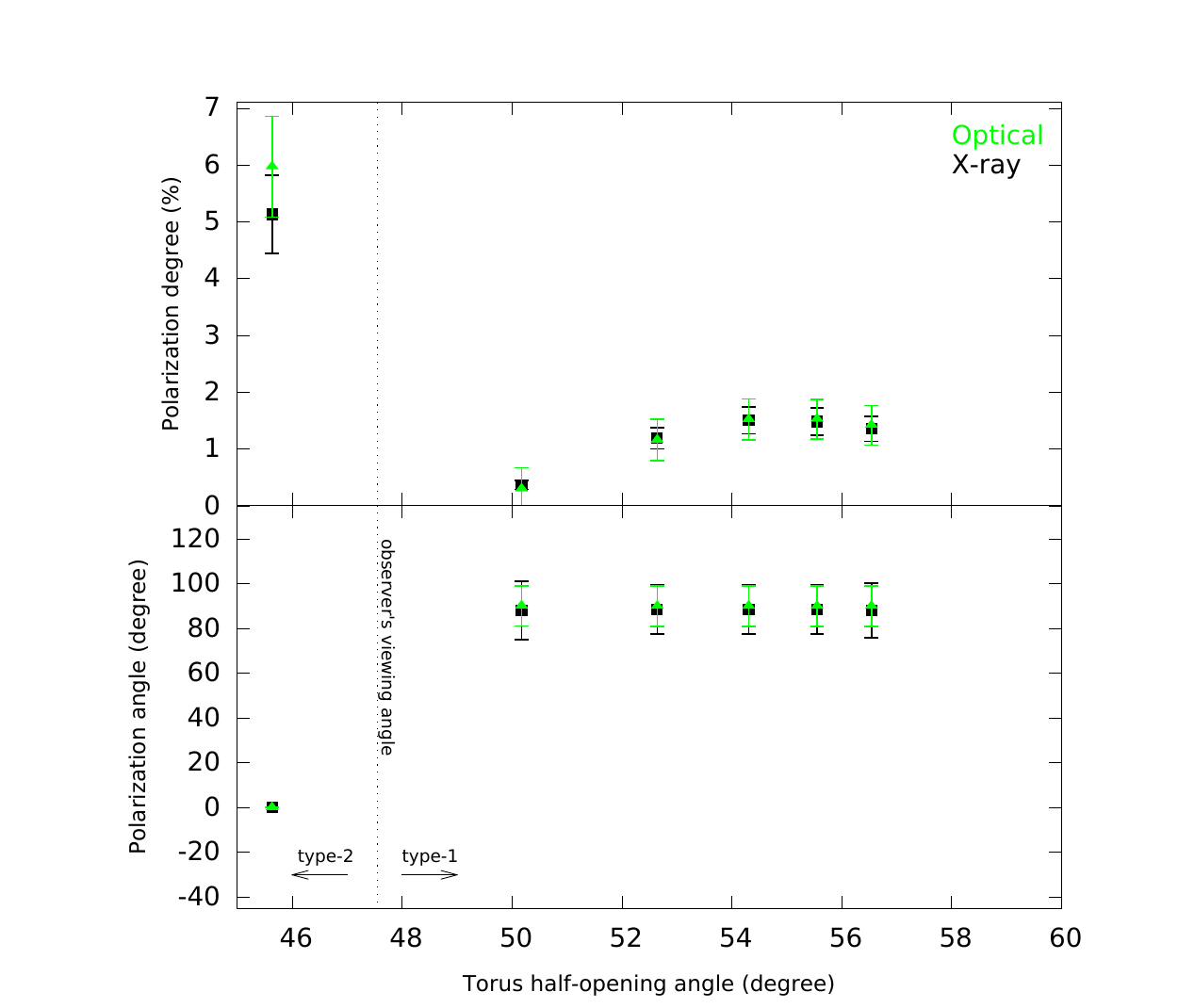}
      \caption{Computed optical (green) and X-ray 
               (black) polarization as a function of 
               the torus half-opening angle (defined with 
               respect to the symmetry axis of the model) 
               in the framework of the receding torus 
               scheme. The line of sight (dashed line) 
               and AGN geometry correspond to those 
               of NGC~4151.}
     \label{Fig:X_variation}%
   \end{figure}

The second NGC~4151 scenario allows the torus morphology to vary instantaneously and the parametrization of 
the model follows the calculations presented in Sect.~\ref{Model}. The output of {\sc stokes} is shown in 
Fig.~\ref{Fig:NGC_4151_variation}. In the case of the receding torus scheme, both the degree and angle of 
polarization (either in the optical and soft X-ray band) are sensitive to the variation in flux. 
However, as illustrated in Fig.~\ref{Fig:X_variation}, the $P$- and $PPA$-dependences are not linearly 
correlated with the half-opening angle of the torus (linked with $\Gamma$, see Fig.~\ref{Fig:Parametrization}). 
The net polarization strongly depends on the radiative coupling between the different AGN constituents where 
the contribution from the polar outflows strongly contributes to the global polarization properties of NGC~4151. 
For hard $\Gamma$ indexes ($\le$ 1.4), the inner radius of the torus is closer to the central black hole 
(R$_{\rm in}$ = 0.04~pc) and the obscuring equatorial torus is puffed up by the lack of intense radiation 
coming from the central engine (half-opening angle of 45.6$^\circ$). The observer's line of sight becomes 
obstructed by gas and dust clouds that prevent a direct view of the torus funnel. Radiation mainly
escapes by polar scattering onto the ionized winds, giving rise to strong ($\sim$ 4\% -- 6\%) perpendicular 
polarization. However, at $\Gamma$ = 1.5 the torus horizon (half-opening angle of 50.2$^\circ$) no longer 
intercepts the line of sight of the distant observer  and the equatorial scattering contribution starts 
to dominate. The polarization angle switches from perpendicular to parallel and $P$ decreases below 0.4\% owing 
to the canceling contributions of photons from the polar and equatorial scattering structures. Finally, at 
softer $\Gamma$ ($\ge$ 1.6), equatorial scattering dominates and $P$ becomes more stable (1.5\% $\pm$ 0.3\%).

In the framework of the receding torus model, NGC~4151 would be subject to drastic changes in polarization at 
low $\Gamma$: $P$ varies between 0.1\% to 6\% and the polarization position angle can rotate by 90$^\circ$ 
between two epochs and in two different energy bands. If the timescales for the radiation source 
and torus morphology variations are relatively similar, strong optical and X-ray polarimetric signatures 
are thus predicted in connection with the dynamical evolution of the dusty medium in the inner parsec of an AGN.


\section{Discussion}
\label{Discussion}

\subsection{Timescale of polarization variability}
\label{Discussion:Timescales}

What are the timescales for the torus to react to the change in the intensity of the central engine?
The light travel time for a photon emitted close to the central supermassive black hole to reach the 
inner 0.1~pc radius of the torus is about 10$^7$ seconds ($\sim$ 4 months). The dust grains will 
then react to the stronger/weaker radiation field by sublimation/condensation processes and a 
yet-to-be-determined amount of time will be necessary for the torus to balance  its temperature again. 
Constraining the properties of the subparsec dusty environment using variability is not straightforward, 
but several attempts have been made, specifically on NGC~4151. \citet{Koshida2009} tried to estimate 
the timescale for variation in the inner radius of the NGC~4151 torus by cross-correlating its longterm 
(2001 - 2006) optical and NIR emission. The different time lags between the emission from the disk (optical) 
and the torus (IR) were not consistent with a fixed inner radius. In particular, the time lags of 
\citet{Koshida2009} deviate from the $\propto L^\frac{1}{2}$ relation inferred from dust reverberation 
studies \citep{Barvainis1987,Barvainis1992}, pointing towards a morphological variation of one of the 
two emitters. The timescale for dust replenishment in the central region of NGC~4151 was then estimated 
to be about one year in order to match the elapsed time between the variable time lags. A 
second and independent estimation of the timescale for dust formation and destruction in NGC~4151 was 
achieved by \citet{Kishimoto2013} using near-IR interferometry. The authors found that the apparent 
variation of visibility in the observations taken between 2008 and 2011 points toward a slightly longer
delay of a few years to explain the torus inner radius evolution. To narrow down the uncertainties on
timescales, \citet{Honig2011} used a new simulation code to model time-variable IR emission from dusty 
tori and examined the light curves of NGC~4151 extracted from \citet{Koshida2009}. They successfully 
reproduced the observed variability signal using a simplified clumpy environment, and they also 
found that almost half of the energy in the optical variability is converted into IR variability. However, 
the code used by the authors does not yet include a change in sublimation radius with the source intensity. 
Forthcoming realizations of the code with upgraded physics will be able to precisely confirm or infer the one-year timescale estimation made by \citet{Koshida2009}.

If we consider that a year is necessary for the torus to be in temperature equilibrium after dust 
sublimation or condensation resulting from a year-long, stable, bright or quiescent activity, the 
polarimetric variability should lag behind the flux fluctuation by the same amount of time. Therefore, it is useful 
to conduct long-term optical and X-ray polarimetric observations, regularly spaced in one-year periods of time, in order to catch the change in torus morphology with time in the receding torus 
framework. Additionally, for AGN with the same inclination but different time-averaged $\Gamma$, single 
pointing observations will be able to reveal differences in optical and X-ray polarization 
between soft ($\sim$ 2.0) and hard ($\le$ 1.4) photon indexes.

Another important aspect is that short-term variations in the intensity of the central engine
(on timescales of less than 100~ks; see \citealt{Markowitz2003}) will not affect the polarization
measurements. The torus will not have time to react to the  flickering of the 
corona emission (which lasts  less than a day), meaning that a megasecond-long polarimetric observation  sure to be robust
whatever the sudden onset or offset of intensity.

\subsection{Detectability with modern polarimeters}
\label{Discussion:Polarimeters}

Modern X-ray polarimeters are based on the photoelectric effect, a promising and highly sensitive
basis for new polarimeters. There are two concepts, one based on the time-projection chamber technique 
\citep{Black2007} and one on the gas-pixel technology \citep{Costa2001,Bellazzini2010}; the latter has the 
advantage of recovering the two-dimensional position of the incoming photon, allowing for polarization 
mapping. The gas-pixel technology is used in two X-ray polarimetry missions proposed to NASA and ESA 
for a potential launch within the next decade. The $IXPE$ (Imaging X-ray Polarimetry Explorer - NASA/SMEX 
proposal) and $XIPE$ (X-ray Imaging Polarimetry Explorer - ESA M4 candidate) missions should carry a set 
of 2 -- 8~keV polarimeters that are sensitive enough to record the polarization of nearby AGN such as 
NGC~4151 or NGC~1068.

However, X-ray polarization measurements are photon hungry, as the I, Q, and U parameters of the Stokes 
vector (used to quantify linear polarization) must be recorded with sufficient statistics. In the case of an 
X-ray polarimeter with an ideal response to polarization, 4 $\times$ 10$^6$ photons will be needed to detect 
a 1\% polarization signal against the statistical fluctuations, if we accept a 10~$\sigma$ standard deviation. 
Bright AGN only radiate at a rate of a few tens of counts per second per square meter in the X-ray band, so 
X-ray polarimetry is time consuming. Unfortunately, most of AGN situated in the local Universe are fainter
than 10 cts/sec/m$^2$ (and the brightest $z >$ 1 quasars only radiate at a rate of 1~ct/s/m$^2$), so 
exposition times will increase. 

It is then necessary to estimate the amount of time necessary to detect a minimum polarization threshold that
 corresponds to our predicted polarization degrees.In the photon energy range from 2 to 10~keV,
NGC~4151 has a time-averaged flux of 11.3 $\times$ 10$^{-11}$ erg/s/cm$^2$ (equivalent to 5.65~mCrab, 
\citealt{Wang2010}). To detect a polarization signal at the 3$\sigma$ confidence level when the background flux 
is negligible with respect to the source flux, using a total collecting area $>$ 1100~cm$^2$ at 
3~keV,  940~ks are required in the case of a threshold polarization value of 1.21\% (Sect.~\ref{NGC4151:Stability}), 
610~ks when $P$ is 1.50\% (Sect.~\ref{NGC4151:Variation}, soft $\Gamma$), and 52~ks when $P$ is 5.13\% 
(Sect.~\ref{NGC4151:Variation}, hard $\Gamma$). The estimated values of the uncertainties at 1$\sigma$ on 
the polarization angle are lower than 10$^\circ$. Values are estimated using the software developed by the 
$XIPE$ instrumentation team, based on the formulas presented in \citet{Soffitta2013}.


\section{Conclusions and future work}
\label{Conclusions}

We have undertaken broadband simulations of the polarization degree and position angle emerging from scattering 
in a complex AGN environment powered by a flickering continuum source. Our results, spanning from the near-infrared to the X-ray band, allowed us to test our model against optical and IR spectropolarimetric observations
and to predict the amount of polarization a future X-ray polarimeter would detect in NGC~4151. We tested the 
impact of a receding torus, whose inner radius and half-opening angle are dependent on the intensity of the 
radiation field, on the optical and X-ray polarimetric signatures of a prototypical AGN. We critically 
discussed our results in terms of timescales and feasibility with a photoelectric polarimeter. We summarize our 
findings below:

\begin{enumerate}
\item We tested the relevance of our baseline model by reproducing the degree and position angle 
of past NGC~4151 polarimetric observations in the optical and NIR bands. By so doing, we also 
predicted the degree of polarization to be detected in the UV band ($\sim$ 1.5\%).

\item The integrated 2 -- 8~keV polarization in a standard AGN model varies from  a few 
percent at type-1 inclinations  to tens of percent at type-2 viewing angles. Only the viewing angle 
close to the type-1/type-2 transition is affected by the receding torus model. 

\item Applied to NGC~4151, our model predicts that the polarization degree is almost the same in 
the optical and X-ray bands, with an estimated value on the order of 1.21\% $\pm$ 0.34\%. We expect 
a constant parallel polarization position angle if the timescale necessary for the dust to evaporate or 
condense is longer than the long-term flux variability. On the other hand, if the dusty structure is able 
to evolve fast enough, its polarization degree will vary from 0.1\% to 6\% depending on the photon index. 
An orthogonal rotation of the polarization position angle appears at hard $\Gamma$ for both wavebands, 
otherwise the polarization is parallel to the radio axis.

\item The variation of $P$ with long-term flux variability will not be detectable if the torus morphology 
remains static. The same applies to the receding torus framework for soft $\Gamma$. This is a strong argument 
for the reliability of X-ray polarization measurements since variations in $\Gamma$ will not cancel $P$
owing to rapid rotations of the $PPA$.

\item Pointing at changing-look AGN using an X-ray polarimeter, it will be possible to 
constrain the inner radius of the dusty structure, as well as its half-opening angle and the observer's 
inclination thanks to the observed  variability (or lack thereof) in polarization. A joint optical and soft 
X-ray band proposal could greatly strengthen the polarimetric analyses.

\item Using modern photoelectric polarimeters such as the ones on board  the medium-sized mission 
project $XIPE$, we determined that -- in the most conservative case scenario -- 940~ks observing time will be needed to obtained 
a 3$\sigma$ confidence level polarization measurement of NGC~4151. If the receding torus model is real, 
the observing time decreases to 610~ks for $\Gamma \ge$ 1.6, and 52~ks for $\Gamma \le$ 1.4.

\item The ideal monitoring campaign for an X-ray polarimetric mission equipped with a 2 -- 8~keV
instrument and looking at NGC~4151 to constrain its torus inner radius and half-opening angle is a 
long-term program. A preliminary plan is to achieve 600~ks observations spaced apart in time 
according to the timescale for dust replenishment close to the sublimation radius, i.e., 1 year.
\end{enumerate}

While our model is based on the unified scheme for AGN, several improvements can be envisioned.
First, the impact of fragmentation will be investigated in detail using the code {\sc stokes},
where the reprocessed flux and polarization from a clumpy toroidal structure will be modeled.
Our results will be directly comparable to the recent simulations by \citet{Furui2016} and 
\citet{He2016}, with the advantage of simultaneously predicting the polarization degree and position 
angle of smooth and clumpy molecular tori. This work will be conducted in a wide range of 
wavelengths, from the infrared to the hard X-ray band in order to look in great detail at the 
 receding torus scheme.

We also aim to implement the impact of general relativistic effects close to the central supermassive 
black hole. The thermally emitted UV photons, which travel from the accretion disk to the corona, will
be scattered to higher energy and will then either fall back onto the disk or  travel to the observer.
The trajectories of the photos will follow the photon-null geodesics, at least in the few tens of 
gravitational radii around the potential well. According to \citet{Connors1977} and \citet{Connors1980},
the polarization position angle of radiation will change non-linearly according to the physical parameters 
of the model. Using general relativistic routines \citep{Dovciak2011}, we will investigate to what degree a polarized 
primary source will affect the net polarization detected by  a future X-ray polarimeter.

\begin{acknowledgements}
The authors are grateful to the referee Makoto Kishimoto for his useful and constructive comments 
on the manuscript. This research has been partially funded by the French PNHE. FM and RG were supported by 
the grant ANR-11-JS56-013-01 "POLIOPTIX" and by the European Union Seventh Framework Programme (FP7/2007-2013) 
under grant agreement no. 312789 ``StrongGravity''. POP acknowledges financial support from CNES.
\end{acknowledgements}

\bibliographystyle{aa}
\bibliography{biblio}

\begin{thebibliography}{82}
\expandafter\ifx\csname natexlab\endcsname\relax\def\natexlab#1{#1}\fi

\bibitem[{{Anders} \& {Ebihara}(1982)}]{Anders1982}
{Anders}, E. \& {Ebihara}, M. 1982, \gca, 46, 2363

\bibitem[{{Antonucci}(1993)}]{Antonucci1993}
{Antonucci}, R. 1993, \araa, 31, 473

\bibitem[{{Antonucci} \& {Miller}(1985)}]{Antonucci1985}
{Antonucci}, R.~R.~J. \& {Miller}, J.~S. 1985, \apj, 297, 621

\bibitem[{{Arshakian}(2005)}]{Arshakian2005}
{Arshakian}, T.~G. 2005, \aap, 436, 817

\bibitem[{{Barvainis}(1987)}]{Barvainis1987}
{Barvainis}, R. 1987, \apj, 320, 537

\bibitem[{{Barvainis}(1992)}]{Barvainis1992}
{Barvainis}, R. 1992, \apj, 400, 502

\bibitem[{{Bellazzini} \& {Muleri}(2010)}]{Bellazzini2010}
{Bellazzini}, R. \& {Muleri}, F. 2010, Nuclear Instruments and Methods in
  Physics Research A, 623, 766

\bibitem[{{Bianchi} {et~al.}(2012){Bianchi}, {Panessa}, {Barcons}, {Carrera},
  {La Franca}, {Matt}, {Onori}, {Wolter}, {Corral}, {Monaco}, {Ruiz}, \&
  {Brightman}}]{Bianchi2012}
{Bianchi}, S., {Panessa}, F., {Barcons}, X., {et~al.} 2012, \mnras, 426, 3225

\bibitem[{{Black} {et~al.}(2007){Black}, {Baker}, {Deines-Jones}, {Hill}, \&
  {Jahoda}}]{Black2007}
{Black}, J.~K., {Baker}, R.~G., {Deines-Jones}, P., {Hill}, J.~E., \& {Jahoda},
  K. 2007, Nuclear Instruments and Methods in Physics Research A, 581, 755

\bibitem[{{Cassidy} \& {Raine}(1996)}]{Cassidy1996}
{Cassidy}, I. \& {Raine}, D.~J. 1996, \aap, 310, 49

\bibitem[{{Chiang}(2002)}]{Chiang2002}
{Chiang}, J. 2002, \apj, 572, 79

\bibitem[{{Clavel} {et~al.}(1989){Clavel}, {Wamsteker}, \&
  {Glass}}]{Clavel1989}
{Clavel}, J., {Wamsteker}, W., \& {Glass}, I.~S. 1989, \apj, 337, 236

\bibitem[{{Connors} \& {Stark}(1977)}]{Connors1977}
{Connors}, P.~A. \& {Stark}, R.~F. 1977, \nat, 269, 128

\bibitem[{{Connors} {et~al.}(1980){Connors}, {Stark}, \& {Piran}}]{Connors1980}
{Connors}, P.~A., {Stark}, R.~F., \& {Piran}, T. 1980, \apj, 235, 224

\bibitem[{{Costa} {et~al.}(2001){Costa}, {Soffitta}, {Bellazzini}, {Brez},
  {Lumb}, \& {Spandre}}]{Costa2001}
{Costa}, E., {Soffitta}, P., {Bellazzini}, R., {et~al.} 2001, \nat, 411, 662

\bibitem[{{Das} {et~al.}(2005){Das}, {Crenshaw}, {Hutchings}, {Deo}, {Kraemer},
  {Gull}, {Kaiser}, {Nelson}, \& {Weistrop}}]{Das2005}
{Das}, V., {Crenshaw}, D.~M., {Hutchings}, J.~B., {et~al.} 2005, \aj, 130, 945

\bibitem[{{Dorodnitsyn} {et~al.}(2015){Dorodnitsyn}, {Kallman}, \&
  {Proga}}]{Dorodnitsyn2015}
{Dorodnitsyn}, A., {Kallman}, T., \& {Proga}, D. 2015, ArXiv e-prints

\bibitem[{{Dov{\v c}iak} {et~al.}(2011){Dov{\v c}iak}, {Muleri}, {Goosmann},
  {Karas}, \& {Matt}}]{Dovciak2011}
{Dov{\v c}iak}, M., {Muleri}, F., {Goosmann}, R.~W., {Karas}, V., \& {Matt}, G.
  2011, \apj, 731, 75

\bibitem[{{Draine}(2003)}]{Draine2003}
{Draine}, B.~T. 2003, \apj, 598, 1026

\bibitem[{{Draper} {et~al.}(1992){Draper}, {Gledhill}, {Scarrott}, \&
  {Tadhunter}}]{Draper1992}
{Draper}, P.~W., {Gledhill}, T.~M., {Scarrott}, S.~M., \& {Tadhunter}, C.~N.
  1992, \mnras, 257, 309

\bibitem[{{Elvis}(2000)}]{Elvis2000}
{Elvis}, M. 2000, \apj, 545, 63

\bibitem[{{Elvis} {et~al.}(1994){Elvis}, {Wilkes}, {McDowell}, {Green},
  {Bechtold}, {Willner}, {Oey}, {Polomski}, \& {Cutri}}]{Elvis1994}
{Elvis}, M., {Wilkes}, B.~J., {McDowell}, J.~C., {et~al.} 1994, \apjs, 95, 1

\bibitem[{{Fiore} {et~al.}(1998){Fiore}, {Elvis}, {Giommi}, \&
  {Padovani}}]{Fiore1998}
{Fiore}, F., {Elvis}, M., {Giommi}, P., \& {Padovani}, P. 1998, \apj, 492, 79

\bibitem[{{Fischer} {et~al.}(2013){Fischer}, {Crenshaw}, {Kraemer}, \&
  {Schmitt}}]{Fischer2013}
{Fischer}, T.~C., {Crenshaw}, D.~M., {Kraemer}, S.~B., \& {Schmitt}, H.~R.
  2013, \apjs, 209, 1

\bibitem[{{Furui} {et~al.}(2016){Furui}, {Fukazawa}, {Odaka}, {Kawaguchi},
  {Ohno}, \& {Hayashi}}]{Furui2016}
{Furui}, S., {Fukazawa}, Y., {Odaka}, H., {et~al.} 2016, ArXiv e-prints

\bibitem[{{Goosmann} \& {Gaskell}(2007)}]{Goosmann2007}
{Goosmann}, R.~W. \& {Gaskell}, C.~M. 2007, \aap, 465, 129

\bibitem[{{Goosmann} \& {Matt}(2011)}]{Goosmann2011}
{Goosmann}, R.~W. \& {Matt}, G. 2011, \mnras, 415, 3119

\bibitem[{{Gopal-Krishna} {et~al.}(1996){Gopal-Krishna}, {Kulkarni}, \&
  {Wiita}}]{Gopal1996}
{Gopal-Krishna}, {Kulkarni}, V.~K., \& {Wiita}, P.~J. 1996, \apjl, 463, L1

\bibitem[{{Grimes} {et~al.}(2004){Grimes}, {Rawlings}, \&
  {Willott}}]{Grimes2004}
{Grimes}, J.~A., {Rawlings}, S., \& {Willott}, C.~J. 2004, \mnras, 349, 503

\bibitem[{{Gu} \& {Cao}(2009)}]{Gu2009}
{Gu}, M. \& {Cao}, X. 2009, \mnras, 399, 349

\bibitem[{{He} {et~al.}(2016){He}, {Liu}, \& {Zhang}}]{He2016}
{He}, J.-J., {Liu}, Y., \& {Zhang}, S.-N. 2016, \mnras, 455, 3968

\bibitem[{{Hill} {et~al.}(1996){Hill}, {Goodrich}, \& {Depoy}}]{Hill1996}
{Hill}, G.~J., {Goodrich}, R.~W., \& {Depoy}, D.~L. 1996, \apj, 462, 163

\bibitem[{{Hoffman} \& {Draine}(2016)}]{Hoffman2016}
{Hoffman}, J. \& {Draine}, B.~T. 2016, \apj, 817, 139

\bibitem[{{H{\"o}nig} \& {Kishimoto}(2011)}]{Honig2011}
{H{\"o}nig}, S.~F. \& {Kishimoto}, M. 2011, \aap, 534, A121

\bibitem[{{Jackson} \& {Browne}(1990)}]{Jackson1990}
{Jackson}, N. \& {Browne}, I.~W.~A. 1990, \nat, 343, 43

\bibitem[{{Kapahi} \& {Kulkarni}(1990)}]{Kapahi1990}
{Kapahi}, V.~K. \& {Kulkarni}, V.~K. 1990, \aj, 99, 1397

\bibitem[{{Kemp} {et~al.}(1977){Kemp}, {Rieke}, {Lebofsky}, \&
  {Coyne}}]{Kemp1977}
{Kemp}, J.~C., {Rieke}, G.~H., {Lebofsky}, M.~J., \& {Coyne}, G.~V. 1977,
  \apjl, 215, L107

\bibitem[{{Kishimoto} {et~al.}(2013){Kishimoto}, {H{\"o}nig}, {Antonucci},
  {Millan-Gabet}, {Barvainis}, {Millour}, {Kotani}, {Tristram}, \&
  {Weigelt}}]{Kishimoto2013}
{Kishimoto}, M., {H{\"o}nig}, S.~F., {Antonucci}, R., {et~al.} 2013, \apjl,
  775, L36

\bibitem[{{Kishimoto} {et~al.}(2007){Kishimoto}, {H{\"o}nig}, {Beckert}, \&
  {Weigelt}}]{Kishimoto2007}
{Kishimoto}, M., {H{\"o}nig}, S.~F., {Beckert}, T., \& {Weigelt}, G. 2007,
  \aap, 476, 713

\bibitem[{{Kishimoto} {et~al.}(2009){Kishimoto}, {H{\"o}nig}, {Tristram}, \&
  {Weigelt}}]{Kishimoto2009}
{Kishimoto}, M., {H{\"o}nig}, S.~F., {Tristram}, K.~R.~W., \& {Weigelt}, G.
  2009, \aap, 493, L57

\bibitem[{{Koshida} {et~al.}(2009){Koshida}, {Yoshii}, {Kobayashi}, {Minezaki},
  {Sakata}, {Sugawara}, {Enya}, {Suganuma}, {Tomita}, {Aoki}, \&
  {Peterson}}]{Koshida2009}
{Koshida}, S., {Yoshii}, Y., {Kobayashi}, Y., {et~al.} 2009, \apjl, 700, L109

\bibitem[{{Lawrence}(1991)}]{Lawrence1991}
{Lawrence}, A. 1991, \mnras, 252, 586

\bibitem[{{Marchese} {et~al.}(2012){Marchese}, {Della Ceca}, {Caccianiga},
  {Severgnini}, {Corral}, \& {Fanali}}]{Marchese2012}
{Marchese}, E., {Della Ceca}, R., {Caccianiga}, A., {et~al.} 2012, \aap, 539,
  A48

\bibitem[{{Marin} \& {Goosmann}(2014)}]{Marin2014}
{Marin}, F. \& {Goosmann}, R.~W. 2014, in SF2A-2014: Proceedings of the Annual
  meeting of the French Society of Astronomy and Astrophysics, ed. J.~{Ballet},
  F.~{Martins}, F.~{Bournaud}, R.~{Monier}, \& C.~{Reyl{\'e}}, 103--108

\bibitem[{{Marin} {et~al.}(2015{\natexlab{a}}){Marin}, {Goosmann}, \&
  {Gaskell}}]{Marin2015}
{Marin}, F., {Goosmann}, R.~W., \& {Gaskell}, C.~M. 2015{\natexlab{a}}, \aap,
  577, A66

\bibitem[{{Marin} {et~al.}(2012){Marin}, {Goosmann}, {Gaskell}, {Porquet}, \&
  {Dov{\v c}iak}}]{Marin2012}
{Marin}, F., {Goosmann}, R.~W., {Gaskell}, C.~M., {Porquet}, D., \& {Dov{\v
  c}iak}, M. 2012, \aap, 548, A121

\bibitem[{{Marin} {et~al.}(2015{\natexlab{b}}){Marin}, {Muleri}, {Soffitta},
  {Karas}, \& {Kunneriath}}]{Marin2015b}
{Marin}, F., {Muleri}, F., {Soffitta}, P., {Karas}, V., \& {Kunneriath}, D.
  2015{\natexlab{b}}, \aap, 576, A19

\bibitem[{{Markowitz} {et~al.}(2003){Markowitz}, {Edelson}, \&
  {Vaughan}}]{Markowitz2003}
{Markowitz}, A., {Edelson}, R., \& {Vaughan}, S. 2003, \apj, 598, 935

\bibitem[{{Martel}(1998)}]{Martel1998}
{Martel}, A.~R. 1998, \apj, 508, 657

\bibitem[{{Martin} {et~al.}(1983){Martin}, {Thompson}, {Maza}, \&
  {Angel}}]{Martin1983}
{Martin}, P.~G., {Thompson}, I.~B., {Maza}, J., \& {Angel}, J.~R.~P. 1983,
  \apj, 266, 470

\bibitem[{{Mathis} {et~al.}(1977){Mathis}, {Rumpl}, \&
  {Nordsieck}}]{Mathis1977}
{Mathis}, J.~S., {Rumpl}, W., \& {Nordsieck}, K.~H. 1977, \apj, 217, 425

\bibitem[{{Minezaki} {et~al.}(2004){Minezaki}, {Yoshii}, {Kobayashi}, {Enya},
  {Suganuma}, {Tomita}, {Aoki}, \& {Peterson}}]{Minezaki2004}
{Minezaki}, T., {Yoshii}, Y., {Kobayashi}, Y., {et~al.} 2004, \apjl, 600, L35

\bibitem[{{Mundell} {et~al.}(2003){Mundell}, {Wrobel}, {Pedlar}, \&
  {Gallimore}}]{Mundell2003}
{Mundell}, C.~G., {Wrobel}, J.~M., {Pedlar}, A., \& {Gallimore}, J.~F. 2003,
  \apj, 583, 192

\bibitem[{{Nandra}(2001)}]{Nandra2001}
{Nandra}, K. 2001, Advances in Space Research, 28, 295

\bibitem[{{Oknyanskij} \& {Horne}(2001)}]{Oknyanskij2001}
{Oknyanskij}, V.~L. \& {Horne}, K. 2001, in Astronomical Society of the Pacific
  Conference Series, Vol. 224, Probing the Physics of Active Galactic Nuclei,
  ed. B.~M. {Peterson}, R.~W. {Pogge}, \& R.~S. {Polidan}, 149

\bibitem[{{Panessa} {et~al.}(2009){Panessa}, {Carrera}, {Bianchi}, {Corral},
  {Gastaldello}, {Barcons}, {Bassani}, {Matt}, \& {Monaco}}]{Panessa2009}
{Panessa}, F., {Carrera}, F.~J., {Bianchi}, S., {et~al.} 2009, \mnras, 398,
  1951

\bibitem[{{Penston} {et~al.}(1990){Penston}, {Robinson}, {Alloin},
  {Appenzeller}, {Aretxaga}, {Axon}, {Baribaud}, {Barthel}, {Baum}, {Boisson},
  {de Bruyn}, {Clavel}, {Colina}, {Dennefeld}, {Diaz}, {Dietrich}, {Durret},
  {Dyson}, {Gondhalekar}, {van Groningen}, {Jablonka}, {Jackson},
  {Kollatschny}, {Laurikainen}, {Lawrence}, {Masegosa}, {McHardy}, {Meurs},
  {Miley}, {Moles}, {O'Brien}, {O'Dea}, {del Olmo}, {Pedlar}, {Perea}, {Perez},
  {Perez-Fournon}, {Perry}, {Pilbratt}, {Rees}, {Robson}, {Rodriguez-Pascual},
  {Rodriguez Espinosa}, {Santos-Lleo}, {Schilizzi}, {Stasi{\'n}ska}, {Stirpe},
  {Tadhunter}, {Terlevich}, {Terlevich}, {Unger}, {Vila-Vilaro}, {Vilchez},
  {Wagner}, {Ward}, \& {Yates}}]{Penston1990}
{Penston}, M.~V., {Robinson}, A., {Alloin}, D., {et~al.} 1990, \aap, 236, 53

\bibitem[{{Ponti} {et~al.}(2006){Ponti}, {Miniutti}, {Cappi}, {Maraschi},
  {Fabian}, \& {Iwasawa}}]{Ponti2006}
{Ponti}, G., {Miniutti}, G., {Cappi}, M., {et~al.} 2006, \mnras, 368, 903

\bibitem[{{Proga}(2007)}]{Proga2007}
{Proga}, D. 2007, \apj, 661, 693

\bibitem[{{Puccetti} {et~al.}(2007){Puccetti}, {Fiore}, {Risaliti}, {Capalbi},
  {Elvis}, \& {Nicastro}}]{Puccetti2007}
{Puccetti}, S., {Fiore}, F., {Risaliti}, G., {et~al.} 2007, \mnras, 377, 607

\bibitem[{{Revnivtsev} {et~al.}(2009){Revnivtsev}, {Sazonov}, {Churazov},
  {Forman}, {Vikhlinin}, \& {Sunyaev}}]{Revnivtsev2009}
{Revnivtsev}, M., {Sazonov}, S., {Churazov}, E., {et~al.} 2009, \nat, 458, 1142

\bibitem[{{Risaliti} {et~al.}(2005){Risaliti}, {Elvis}, {Fabbiano}, {Baldi}, \&
  {Zezas}}]{Risaliti2005}
{Risaliti}, G., {Elvis}, M., {Fabbiano}, G., {Baldi}, A., \& {Zezas}, A. 2005,
  \apjl, 623, L93

\bibitem[{{Robinson} {et~al.}(1994){Robinson}, {Vila-Vilaro}, {Axon}, {Perez},
  {Wagner}, {Baum}, {Boisson}, {Durret}, {Gonzalez-Delgado}, {Moles},
  {Masegosa}, {O'Brien}, {O'Dea}, {del Olmo}, {Pedlar}, {Penston}, {Perea},
  {Perez-Fournon}, {Rodriguez-Espinosa}, {Tadhunter}, {Terlevich}, {Unger}, \&
  {Ward}}]{Robinson1994}
{Robinson}, A., {Vila-Vilaro}, B., {Axon}, D.~J., {et~al.} 1994, \aap, 291, 351

\bibitem[{{Ruiz} {et~al.}(2003){Ruiz}, {Young}, {Packham}, {Alexander}, \&
  {Hough}}]{Ruiz2003}
{Ruiz}, M., {Young}, S., {Packham}, C., {Alexander}, D.~M., \& {Hough}, J.~H.
  2003, \mnras, 340, 733

\bibitem[{{Sanders} {et~al.}(1989){Sanders}, {Phinney}, {Neugebauer}, {Soifer},
  \& {Matthews}}]{Sanders1989}
{Sanders}, D.~B., {Phinney}, E.~S., {Neugebauer}, G., {Soifer}, B.~T., \&
  {Matthews}, K. 1989, \apj, 347, 29

\bibitem[{{Schmidt} \& {Miller}(1980)}]{Schmidt1980}
{Schmidt}, G.~D. \& {Miller}, J.~S. 1980, \apj, 240, 759

\bibitem[{{Schn{\"u}lle} {et~al.}(2015){Schn{\"u}lle}, {Pott}, {Rix},
  {Peterson}, {De Rosa}, \& {Shappee}}]{Schnulle2015}
{Schn{\"u}lle}, K., {Pott}, J.-U., {Rix}, H.-W., {et~al.} 2015, \aap, 578, A57

\bibitem[{{Simpson}(2005)}]{Simpson2005}
{Simpson}, C. 2005, \mnras, 360, 565

\bibitem[{{Singal} {et~al.}(2012){Singal}, {Petrosian}, \&
  {Ajello}}]{Singal2012}
{Singal}, J., {Petrosian}, V., \& {Ajello}, M. 2012, \apj, 753, 45

\bibitem[{{Smith} \& {Done}(1996)}]{Smith1996}
{Smith}, D.~A. \& {Done}, C. 1996, \mnras, 280, 355

\bibitem[{{Smith} {et~al.}(2004){Smith}, {Robinson}, {Alexander}, {Young},
  {Axon}, \& {Corbett}}]{Smith2004}
{Smith}, J.~E., {Robinson}, A., {Alexander}, D.~M., {et~al.} 2004, \mnras, 350,
  140

\bibitem[{{Soffitta} {et~al.}(2013){Soffitta}, {Barcons}, {Bellazzini},
  {Braga}, {Costa}, {Fraser}, {Gburek}, {Huovelin}, {Matt}, {Pearce},
  {Poutanen}, {Reglero}, {Santangelo}, {Sunyaev}, {Tagliaferri}, {Weisskopf},
  {Aloisio}, {Amato}, {Attin{\'a}}, {Axelsson}, {Baldini}, {Basso}, {Bianchi},
  {Blasi}, {Bregeon}, {Brez}, {Bucciantini}, {Burderi}, {Burwitz}, {Casella},
  {Churazov}, {Civitani}, {Covino}, {Curado da Silva}, {Cusumano}, {Dadina},
  {D'Amico}, {De Rosa}, {Di Cosimo}, {Di Persio}, {Di Salvo}, {Dovciak},
  {Elsner}, {Eyles}, {Fabian}, {Fabiani}, {Feng}, {Giarrusso}, {Goosmann},
  {Grandi}, {Grosso}, {Israel}, {Jackson}, {Kaaret}, {Karas}, {Kuss}, {Lai},
  {Rosa}, {Larsson}, {Larsson}, {Latronico}, {Maggio}, {Maia}, {Marin},
  {Massai}, {Mineo}, {Minuti}, {Moretti}, {Muleri}, {O'Dell}, {Pareschi},
  {Peres}, {Pesce}, {Petrucci}, {Pinchera}, {Porquet}, {Ramsey}, {Rea},
  {Reale}, {Rodrigo}, {R{\'o}{\.z}a{\'n}ska}, {Rubini}, {Rudawy}, {Ryde},
  {Salvati}, {de Santiago}, {Sazonov}, {Sgr{\'o}}, {Silver}, {Spandre},
  {Spiga}, {Stella}, {Tamagawa}, {Tamborra}, {Tavecchio}, {Teixeira Dias}, {van
  Adelsberg}, {Wu}, \& {Zane}}]{Soffitta2013}
{Soffitta}, P., {Barcons}, X., {Bellazzini}, R., {et~al.} 2013, Experimental
  Astronomy, 36, 523

\bibitem[{{Steffen} {et~al.}(2003){Steffen}, {Barger}, {Cowie}, {Mushotzky}, \&
  {Yang}}]{Steffen2003}
{Steffen}, A.~T., {Barger}, A.~J., {Cowie}, L.~L., {Mushotzky}, R.~F., \&
  {Yang}, Y. 2003, \apjl, 596, L23

\bibitem[{{Thompson} {et~al.}(1979){Thompson}, {Angel}, {Stockman}, {Woolf},
  {Martin}, {Maza}, {Beaver}, \& {Landstreet}}]{Thompson1979}
{Thompson}, I., {Angel}, J.~R.~P., {Stockman}, H.~S., {et~al.} 1979, \apj, 229,
  909

\bibitem[{{Turner} {et~al.}(1997){Turner}, {George}, {Nandra}, \&
  {Mushotzky}}]{Turner1997}
{Turner}, T.~J., {George}, I.~M., {Nandra}, K., \& {Mushotzky}, R.~F. 1997,
  \apjs, 113, 23

\bibitem[{{Urry} \& {Padovani}(1995)}]{Urry1995}
{Urry}, C.~M. \& {Padovani}, P. 1995, \pasp, 107, 803

\bibitem[{{Ursini} {et~al.}(2015){Ursini}, {Boissay}, {Petrucci}, {Matt},
  {Cappi}, {Bianchi}, {Kaastra}, {Harrison}, {Walton}, {di Gesu}, {Costantini},
  {De Marco}, {Kriss}, {Mehdipour}, {Paltani}, {Peterson}, {Ponti}, \&
  {Steenbrugge}}]{Ursini2015}
{Ursini}, F., {Boissay}, R., {Petrucci}, P.-O., {et~al.} 2015, \aap, 577, A38

\bibitem[{{Wang} {et~al.}(2010){Wang}, {Risaliti}, {Fabbiano}, {Elvis},
  {Zezas}, \& {Karovska}}]{Wang2010}
{Wang}, J., {Risaliti}, G., {Fabbiano}, G., {et~al.} 2010, \apj, 714, 1497

\bibitem[{{Weymann} {et~al.}(1991){Weymann}, {Morris}, {Foltz}, \&
  {Hewett}}]{Weymann1991}
{Weymann}, R.~J., {Morris}, S.~L., {Foltz}, C.~B., \& {Hewett}, P.~C. 1991,
  \apj, 373, 23

\bibitem[{{Willott} {et~al.}(2000){Willott}, {Rawlings}, {Blundell}, \&
  {Lacy}}]{Willott2000}
{Willott}, C.~J., {Rawlings}, S., {Blundell}, K.~M., \& {Lacy}, M. 2000,
  \mnras, 316, 449

\bibitem[{{Willott} {et~al.}(2001){Willott}, {Rawlings}, {Blundell}, {Lacy}, \&
  {Eales}}]{Willott2001}
{Willott}, C.~J., {Rawlings}, S., {Blundell}, K.~M., {Lacy}, M., \& {Eales},
  S.~A. 2001, \mnras, 322, 536

\bibitem[{{Young}(2000)}]{Young2000}
{Young}, S. 2000, \mnras, 312, 567

\end{thebibliography}

\end{document}